\documentclass[reprint,aps,prl,superscriptaddress,longbibliography]{revtex4-2}

\usepackage{amsmath, amssymb, times, mathrsfs, hyperref, array, bbm}
\usepackage{graphicx}
\usepackage[usenames,dvipsnames]{xcolor}
\usepackage{braket}
\usepackage{xcolor}
\usepackage{tcolorbox}
\usepackage{listings}
\hypersetup{colorlinks=false}
\usepackage{hyperref}

\bibpunct{[}{]}{,}{n}{}{}
\usepackage{ulem}
\usepackage{lipsum}
\usepackage{listings}

\usepackage{mdframed}

\begin{document}
\title{Comment on ``InAs-Al hybrid devices passing the topological gap protocol",\\
Microsoft Quantum, Phys. Rev. B 107, 245423 (2023)}

\author{Henry F. Legg}
\affiliation{SUPA, School of Physics and Astronomy, University of St Andrews, North Haugh, St Andrews, KY16 9SS, United Kingdom}
\affiliation{Department of Physics, University of Basel, Klingelbergstrasse 82, CH-4056 Basel, Switzerland}

\begin{abstract}
The topological gap protocol (TGP) is presented as “a series of stringent experimental tests" for the presence of topological superconductivity and associated Majorana zero modes. Here, we show that the TGP, ``passed" by Microsoft Quantum [PRB 107, 245423 (2023)], lacks a consistent definition of  ``gap'' or ``topological'', and even utilises different code when applied to theoretical simulations compared to experimental data. Furthermore, the outcome of the TGP is sensitive to the choice of magnetic field range, bias voltage range, data resolution, and cutter voltage pair (junction transparency) --- data parameters that, in PRB~107,~245423~(2023), vary significantly, even for measurements of the same device. As a result, the core claims of PRB~107,~245423~(2023) are based on unexplained measurement choices and inconsistent definitions. In particular, this means the claim by Microsoft Quantum in PRB~107,~245423~(2023) that their devices have a “high probability of being in the topological phase” is not reliable and should be revisited. In light of this, subsequent studies that are based on tuning up devices via the TGP should also be revisited.
\end{abstract}
\maketitle
\onecolumngrid
\twocolumngrid

The search for topological superconductivity and associated {Majorana zero modes (MZMs)} has drawn considerable interest over the last decade, largely due to the potential application of {MZMs} as topological qubits. However, reliably identifying {MZMs} has been an ongoing challenge. This is because {nontopological} effects --- {such as trivial Andreev bound states~\cite{Andreev1964Thermal,Andreev1966Electron,Prada2020From,valentini2020nontopological} in quantum dots~\cite{Lee2012Zero,Liu2017Andreev,reeg2018zero, Cayao2015SNS,Ptok2017Controlling,Liu2019Conductance,Awoga2019Supercurrent,Juenger2020Magnetic}, disorder-induced quasi-Majoranas \cite{kells2012Near,moore2018two,Woods2019Zero, Liu2019Conductance,Vuik2019Reproducing,Penaranda2018Quantifying,Alspaugh2020Volkov,Marra2022Majorana}, and overlapping subgap states~\cite{Chen2019Ubiquitous} --- can mimic the expected signatures of a topological superconducting phase. Recently it has also been shown that nonlocal signatures of a topological superconducting phase transition can be mimicked by disorder effects~\cite{Roshni2023,Hess2023,Loo2023,Legg2024}.

To address {the problems with identifying MZMs}, Microsoft Quantum proposed a ``stringent''~\cite{Agahee2023} and ``unbiased''~\cite{Pikulin2021} test: the so-called \textit{topological gap protocol} (TGP). This protocol combines local and nonlocal conductance data from nanowire devices, which is then processed by ``data analysis routines that allow for an automated and unbiased execution''~\cite{Pikulin2021}. According to Microsoft Quantum, the TGP provides: ``a binary answer to a binary question: is there a topological phase present in the (real or simulated) device that produced this transport data set?''~\cite{Antipov2023}.

In this Comment --- using the publicly released data and code~\cite{Github} for the TGP from Ref.~\citenum{Agahee2023} --- we show that whether the TGP identifies a topological phase in a given dataset depends on the choice of data parameters. Importantly, there are large variations in these data parameters for the measurements reported in Ref.~\citenum{Agahee2023}, even for different measurements on the same device. These issues are compounded by differences between Ref.~\citenum{Agahee2023} and the released code, selective presentation and redefinitions of the TGP throughout {Ref.~\citenum{Agahee2023}}, and differences between the TGP as applied to experiments compared to that applied to theory simulations.

{\bf Issue 1: Identification of the \textit{gap} differs between published paper and released code.} The reliability of the TGP hinges on the accurate identification of the bulk gap of the nanowire. To detect this gap, the TGP utilises a threshold nonlocal conductance, $G_{\rm th}$, such that if the antisymmetrised nonlocal conductance is below this threshold, the TGP treats it as an effective zero conductance. To establish this definition of zero conductance, Ref.~\citenum{Agahee2023} states that $G_{\rm th}$ is set by 5\% ``of the nonlocal conductance at {bias voltages greater than the induced gap}''. That $G_{\rm th}$ is defined above the induced gap is further emphasised by a statement in Ref.~\citenum{Agahee2023} that $G_{\rm th}$ is set by conductance at ``high-bias''. However, an examination of the released code reveals that $G_{\rm th}$ is actually set  using the maximum nonlocal conductance at {\it any} measured bias voltage, including below the induced gap. In fact, most of the measurements in Ref.~\citenum{Agahee2023} do not have a bias voltage range that goes above the induced gap (as defined in Ref.~\citenum{Agahee2023}). Setting this threshold based on the maximum conductance at any bias means that the detection of a ``gap'' can be sensitive to the chosen measurement ranges. For instance, changing the bias voltage range or the magnetic field range can alter the maximum conductance captured in the dataset, shifting $G_{\rm th}$ and flipping the ``binary'' outcome of the TGP.

\begin{figure*}[t]
  \includegraphics[width=1\textwidth]{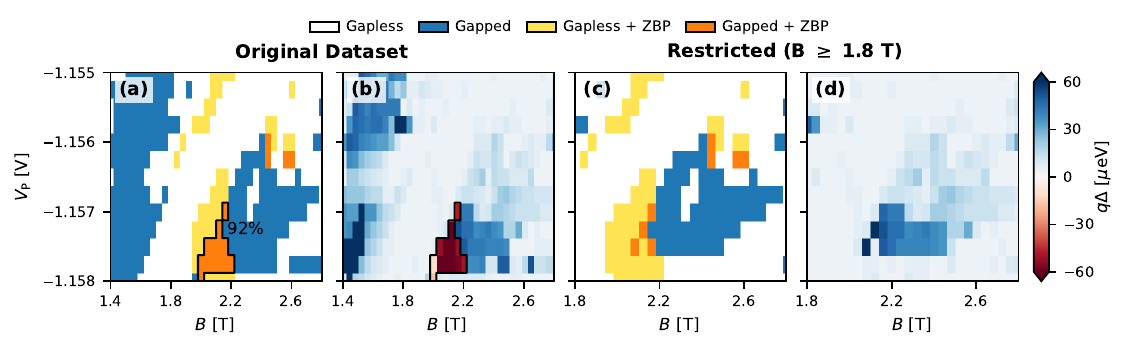}
  \vspace*{-15pt}
	\caption{{\bf Sensitivity of TGP outcome to magnetic field range:} {\bf (a-b)} TGP outcome for Device B (Fig.~16 of Ref.~\citenum{Agahee2023}) with supplied magnetic field range 1.4~T--2.8~T. This passes the TGP, with the orange region around $B\approx 2$~T identified as topological. {\bf (c-d)} When the magnetic field range is reduced to 1.8~T--2.8~T, Device B now fails the TGP, even though the identified region around $B\approx 2$~T remains within the new data window. This sensitivity to data ranges arises because the TGP outcome is determined by the maximum nonlocal conductance within a given cut of bias voltage and magnetic field. Since this maximum nonlocal conductance --- and therefore the TGP result --- depends on the selected data window, changing either the magnetic field or bias voltage range can alter the outcome.}
	\vspace*{-15pt}
\label{DeviceB}
\end{figure*}

{\bf Issue 2: Unexplained variations in experimental data parameters that change TGP outcome.} The topology of a device---an intrinsic property---should not depend on measurement choices. Yet, because the TGP relies on the maximum conductance within a given dataset, it can be acutely sensitive to data ranges and other data parameters, such as resolution and cutter pair (junction transparency). Importantly, whilst some small variations of such data parameters might be expected between different measurements, in Ref.~\citenum{Agahee2023} the {\bf a)} magnetic field ranges, {\bf b)} bias voltage ranges, {\bf c)} data resolution, and {\bf d)} number of cutter pairs (junction transparencies) all vary significantly in the released datasets, for reasons that are not explained. These variations can be up to an order of magnitude and there are large differences in these data parameters even for measurements of the same device. 

To illustrate this dependence on data parameters, we focus here on the TGP's sensitivity to the chosen magnetic field range (see Fig.~\ref{DeviceB}). To demonstrate this we use the data of Ref.~\citenum{Agahee2023} provided for Device B, see Fig.~\ref{DeviceB}. As in Ref.~\citenum{Agahee2023}, this device passes the TGP with a magnetic field range of 1.4~T~--~2.8~T with a region around $B\approx 2$ T identified as purportedly topological. However, by reducing the magnetic field range to 1.8~T~--~2.8~T, this device now fails the TGP, even though the purported topological region is still well within this selected data range. The TGP outcome is altered because there is a shift in the maximum in nonlocal conductance due to the change in range, which in turn completely flips the outcome of the protocol.

Most importantly, as shown in Fig.~\ref{rangesB}, the provided datasets for Ref.~\citenum{Agahee2023} show considerable variations in magnetic field ranges, even for the same device (see A1–3). For instance, in Device E the range is 0.4--0.8 T, but for Device A1 the range is 0.5--2.5 T, i.e., the range is 5 times larger. The start and end points of the ranges also vary significantly. Whilst it is explained in Ref.~\citenum{Agahee2023} that the outcome of Stage 1 is used to ``restrict ($B$,$V_p$) to smaller ranges", the way this is chosen is not explained, no similar reduction in range occurs for the Stage 2 data of the simulations (see discussion below), and the released TGP code provides no further clarification. 

\begin{figure}[b]
  \includegraphics[width=0.987\columnwidth]{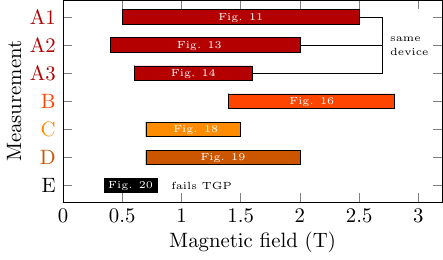}
	\caption{{\bf Magnetic field ranges utilised for devices in Ref.~\citenum{Agahee2023}:} The magnetic field ranges of the measurements reported in Ref.~\citenum{Agahee2023} exhibit considerable and unexplained variations, even for the same device. Since the TGP outcome depends on the nonlocal conductance maximum, altering it can change whether a device `passes' or `fails'.}
\label{rangesB}
\end{figure}

It should be strongly emphasised that it is not just the magnetic field range that can alter the outcome of the TGP: A detailed discussion demonstrating similar dependence on bias voltage range, data resolution, and number of cutter pair voltages (junction transparency) is shown in Appendix B. Most importantly, as with magnetic field range, there are also considerable variations in these data parameters for data released for Ref.~\citenum{Agahee2023}. Namely, there is a factor of 2.5 difference in bias voltage ranges, a large difference in resolutions, and the number of cutter pairs varies from 1 to 5. There are no clear explanations for these variations within Ref.~\citenum{Agahee2023}. 

Finally, we note that the choice of individual cutter pair can alter whether a region is identified as topological or not by the TGP, again due to the dependence on the maximum in conductance. This is despite a statement in Ref.~\citenum{Agahee2023} that identified regions ``corresponding to the different cutter pairs are similar." In fact, this statement is only correct for Device A1 that is presented in the Appendix of Ref.~\citenum{Agahee2023}, but is incorrect for all other devices with multiple cutter pairs. This dependence on individual cutter pair voltages is discussed in further detail in Appendix B3.

The sensitivity of the TGP to data parameters demonstrates that it produces results that can be dictated by measurement choices rather than an underlying property of the studied devices. Most importantly, not only does the TGP have a sensitivity to measurement parameter choices, but there are large variations in the data parameters of the experimental datasets underlying Ref.~\citenum{Agahee2023}. These variations also raise the question whether there are more datasets for these devices with different data parameters that were not reported.

{\bf Issue 3: The TGP for theoretical simulations is not the same as that applied to experimental data.} We now turn from the experimental findings reported in Ref.~\citenum{Agahee2023} to the theoretical underpinnings of the TGP. The core results of Ref.~\citenum{Agahee2023} rely on the claim that the protocol was tested against extensive simulations that yielded ``no false positives'', which leads to the assertion in Table~II of Ref.~\citenum{Agahee2023} that there is a less than $8\%$ probability of a device passing the TGP whilst being in a trivial phase. However, applying the publicly released TGP code to the available simulation data reveals that the claim of no false positives is not correct for the TGP as applied to the experimental data and figures in Ref.~\citenum{Agahee2023}. Genuine false positives do occur in simulations (see Appendix~C for an example) contradicting Ref.~\citenum{Agahee2023}.

This discrepancy arises because the Python code used to establish the claim of ``no false positives'' for Table~II utilises a fundamentally different TGP implementation (\texttt{analyze\_2}) than the one applied to the experimental data and paper figures (\texttt{analyze\_two}). Crucially, these versions differ in their parameter choices, such as whether local conductance is averaged over different cutter voltage pairs (\texttt{average\_over\_cutter} is set to \texttt{False} for \texttt{analyze\_2} but \texttt{True} for \texttt{analyze\_two}) and this leads to different TGP outcomes. 

Beyond these coding inconsistencies, as noted above, there are also significant differences in the data parameters used for the simulations compared to the experiments. For instance, the simulations all contain five cutter pairs whilst only a single cutter pair is used for Device D. Ultimately, these discrepancies in both the underlying code and the chosen data parameters demonstrate that the TGP lacks a consistent definition even within Ref.~\citenum{Agahee2023}, rendering any comparison between theory and experiment fundamentally unsound. Ultimately this means that the ``high probability'' of a topological phase claimed in Ref.~\citenum{Agahee2023} is unreliable due to flawed comparisons between simulations and experiments.

{\bf Issue 4: Redefinition of \textit{topological}.} Finally, continuing our focus on the theoretical aspects of the TGP, we consider the definition of a \textit{true positive} for the theoretical simulations. The original TGP of Pikulin {\it et al.} defined a topological pixel in phase space using a scattering invariant threshold, specifically $\det(r) < -0.9$~\cite{Pikulin2021}. However, Ref.~\citenum{Agahee2023} considerably weakens this requirement, classifying a pixel as topological if $\det(r) < 0$ at {\sl either} nanowire end and for {\sl any} cutter pair. This considerably relaxed definition results in 37.7\% of the entire simulated phase space in Stage 2 simulations being categorised as topological (see Appendix D). 

Furthermore, for an identified region to be deemed a \textit{true positive}, only a single pixel within the region must meet this weakened criterion for topology. This definition of true positive also allows mixed trivial and topological regions to be treated as a single phase, contradicting the protocol's stated core purpose of detecting the transition between the topological and trivial phases.

Finally, we note that the weakness of this redefinition of topology was obscured in Ref.~\citenum{Agahee2023} by selective data presentation. For instance, the classification of pixels as topological is omitted from the only simulation that fails the TGP (Fig.~32 of Ref.~\citenum{Agahee2023}); reinserting them (see Appendix D) reveals that almost all of its phase space in this figure is classified as topological under the weakened criteria. This failing simulation was the only presented simulation where the topological classification of pixels was omitted.

{\bf Conclusions.} Distinguishing trivial from topological states remains a notoriously difficult challenge, especially in the search for MZMs. As such, defining a “stringent” and “binary” test for topological superconductivity was ambitious from the outset. Here we demonstrated how the topological gap protocol falls short of this goal. Not only is the TGP narrowly tailored~\cite{Legg2024} to the specific experiments reported in Ref.~\citenum{Agahee2023} --- and thus lacking broader applicability --- but it is also dependent on data parameter choices, which vary throughout Ref.~\citenum{Agahee2023}. Ultimately, the sensitivity of the TGP to these data parameters largely stems from nonlocal conductance being a poor measure of the bulk band gap~\cite{Hess2023,Loo2023,Legg2026}. Moreover, the unexplained choices of data parameters in Ref.~\citenum{Agahee2023} raise fundamental questions about why these specific  parameters were selected and whether alternative datasets exist for these devices. 

Furthermore, we demonstrated that the TGP is not even consistently defined within Ref.~\citenum{Agahee2023} itself. This means, in particular, the claim in Ref.~\citenum{Agahee2023}: ``Our main result is that several devices... have passed the topological gap protocol defined in Pikulin {\it et al.} (arXiv:2103.12217)'' is not correct. The TGP in Ref.~\citenum{Agahee2023} differs considerably from Pikulin {\it et al.} and within Ref.~\citenum{Agahee2023} itself.

Compounding all this is selective presentation of results, notably the role of cutter pair dependencies and the omission of topological classification of pixels. 

In summary, the lack of a clearly defined TGP and the variations in data parameters cast serious doubt on the reliability of the claim that there is a ``high probability” of topological superconductivity in the devices studied in Ref.~\citenum{Agahee2023} and, by extension, on later studies that rely on the same TGP to tune up devices~\cite{Aghaee2025,Aghaee2025a,Aghaee2026}. These claims should be revisited.

{\bf Code and Data Availability:} No new data was generated in this study. This Comment uses the same TGP code as Ref.~\citenum{Agahee2023}. The figures can be reproduced using the code found on \href{https://doi.org/10.5281/zenodo.21264309}{Zenodo DOI: 10.5281/zenodo.21264309}.
\bibliography{refs}

\appendix

\twocolumngrid

\vspace{-12pt}

\twocolumngrid

\section{Appendix A: Further discussion of identification of the \textit{gap}.} As in the main text, to detect a gap the TGP uses the nonlocal conductance. If the antisymmetrised nonlocal conductance is below the threshold value, $A(G_{\mathrm{RL}}) < G_{\rm th}$, then the TGP treats this as an effective zero conductance. If there is an effective zero nonlocal conductance below 10~$\mu$V, then the TGP detects a gap, otherwise it reports a gapless system. The quantity $G_{\rm t h}$ is therefore, arguably, the most important in the TGP since it determines whether a gap is reported by the topological \textit{gap} protocol. 

It should first be noted that in Ref.~\citenum{Agahee2023}, the antisymmetrised nonlocal conductance is defined as  [Eq.~(D1) of Ref.~\citenum{Agahee2023}]
\begin{equation}
A[G_{\mathrm{RL}}(V_b)] \equiv [G_{\mathrm{RL}}(V_b) - G_{\mathrm{RL}}(-V_b)]/2,\label{antisym}
\end{equation}
where $V_b$ is the bias voltage and $G_{\mathrm{RL}}$ the right-left nonlocal conductance (equivalently $G_{\mathrm{LR}}$ the left-right nonlocal conductance). However, when introducing the need for  $G_{\rm th}$, it is stated: {``$A(G_{\mathrm{RL}})$ and $A(G_{\mathrm{LR}})$ { will never truly vanish at zero bias}."}
As can be seen from  Eq.~\eqref{antisym}, this claim is mathematically incorrect as antisymmetrisation ensures that $A(G_{\mathrm{RL}})$ and $A(G_{\mathrm{LR}})$ exactly vanish at zero bias ($V_b = 0$). 

More fundamentally, the procedure for determining $G_{\rm th}$ in the publicly released code contradicts the published manuscript. The manuscript states:
\begin{quote}
``For the disorder strengths expected in our devices, we take $G_{\rm th}$ equal to $\exp(-3) \approx 0.05$ times the maximal value $\max\{G_{\mathrm{NL}}\}$ of the nonlocal conductance at bias voltages greater than the induced gap...''
\end{quote}
It further asserts that $G_{\rm th}$ is defined in terms of the ``high-bias'' conductance. Moreover, Ref.~\citenum{Agahee2023} explicitly defines this induced gap as ``the superconducting gap that is induced in the wire at $B=0$''. However, in most of the reported experiments, the measured bias voltage ranges are significantly smaller than this. Therefore, setting the threshold based on conductance strictly above the induced gap is not possible for the provided datasets. 

Furthermore, the released TGP code does not implement the manuscript's definition. Instead, the code calculates $G_{\rm th}$ using the maximum nonlocal conductance at \textit{any} measured bias voltage, including low-bias regions. In practice, the maximum nonlocal conductance does frequently occur at very low bias, such as in Device B. Since the code evaluates the maximum across the entire chosen measurement window rather than specifically at high-bias, the resulting threshold --- and therefore the binary outcome of the TGP --- becomes sensitive to the arbitrary choice of measurement data ranges, as discussed in the main text.

\vspace{-7pt}
\section{Appendix B: Further discussion of unexplained variations in experimental data parameters that change TGP outcome.} 
As discussed in the main text, the magnetic field range, bias voltage range, data resolution, and number of cutter pairs (junction transparencies) can all alter the TGP outcome due to the above sensitivity to the maximum in conductance. Here we further  discuss this sensitivity and demonstrate that there are large and unexplained variations throughout Ref.~\citenum{Agahee2023} in several data parameters that can change the TGP outcome.

\twocolumngrid

\onecolumngrid
\vspace{16pt}
{\small
\begin{tcolorbox}[colback=gray!20, colframe=black, sharp corners, boxrule=0.5pt, width=\textwidth]
\begin{verbatim}
223:	gap_threshold_factor: float = 0.05,
224:	upper_conductance_threshold: float = float("inf"),
---------------------------
296:	def gap_thresholding(G: np.ndarray) -> np.ndarray:
297:    f = gap_threshold_factor * min(np.max(G), upper_conductance_threshold)
\end{verbatim}
\end{tcolorbox}}
\noindent {\bf Code for conductance threshold:} A code excerpt from the TGP second stage analysis (\href{https://github.com/microsoft/azure-quantum-tgp/blob/main/tgp/two.py}{two.py}) shows that the function \texttt{gap\_thresholding} uses 
  \texttt{np.max(G)}, {i.e.}, it sets the conductance threshold $G_{\rm th}$ using the maximum over all bias values. This determines when a system is determined to be `gapped' by the TGP, but contradicts the definition set out in Ref.~\citenum{Agahee2023}. 
\vspace{0pt}
 \twocolumngrid

{\bf (B1) Bias voltage range: Dependence and variations.}
In the main text we focussed on magnetic field. However, the TGP is also sensitive to bias voltage range. If the maximum conductance occurs at high-bias then reducing the bias voltage window can alter this maximum, changing $G_{\rm th}$, and altering the TGP outcome. Conversely, extending the bias voltage range can introduce a new nonlocal conductance maximum, again shifting the threshold and altering the TGP outcome. 

\begin{figure}[t]
	\centering
  \includegraphics[width=0.987\columnwidth]{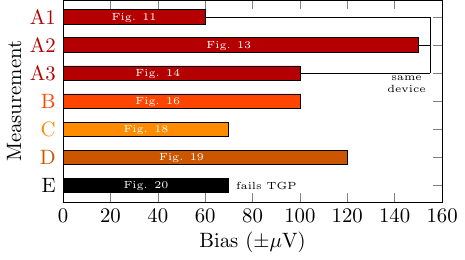}
	\caption{{\bf Bias voltage ranges utilised for devices in Ref.~\citenum{Agahee2023}:} The bias voltage ranges exhibit considerable and unexplained variations, even for the same device. Since the TGP outcome depends on the maximum nonlocal conductance, altering it can change whether a device passes or fails the TGP.  }
\label{rangesBias}
\vspace{-0pt}
\end{figure}

Moreover, since Stage 1 of the TGP detects only zero-bias peaks and not a relevant bias voltage window for gap detection, it is unclear how the selection of bias voltage range can be guided by prior knowledge of the system. Despite this, the released datasets for Ref.~\citenum{Agahee2023} show large and unexplained variations in bias voltage range across different measurements, even for the same device (see Fig.~\ref{rangesBias}). For instance, measurement A1 has a range of $\pm 60$~$\mu$V, while measurement A2 extends over $\pm 150$~$\mu$V, more than 2.5 times larger, even though these are different measurements of the same device. These inconsistencies are not explained in Ref.~\citenum{Agahee2023}, yet they can directly alter the outcome of the TGP. 

Finally we note that, because the bias range determines the maximum possible reported gap, it is somewhat unsurprising that Ref.~\citenum{Agahee2023} reports gaps in the 20–60 $\mu$eV range --- another consequence of measurement choices rather than intrinsic device properties.

\begin{figure}[t]
	\centering
  \includegraphics[width=1\columnwidth]{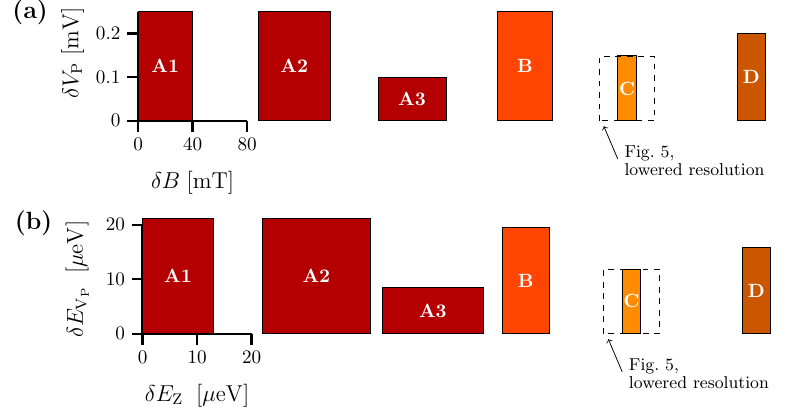}
	\caption{{\bf Variations in resolution (pixel size) of  passing devices in Ref.~\citenum{Agahee2023}:} There are considerable and unexplained variations in resolution across all experiments for both the experimental pixel size {\bf (a)} and the physical pixel size {\bf (b)} (calculated based on the lever arm and g-factors reported in Ref.~\citenum{Agahee2023}). Changing the resolution can affect the TGP outcome: For instance, this is demonstrated for Device C, where reducing the resolution of the experimental data (solid block) to a larger pixel size (dashed line) causes the device to fail the TGP (see Fig.~\ref{resolution}). Notably, the dashed pixel size is still comparable to the resolution used for other devices.}
\label{pxs}
\end{figure}

\begin{figure*}[t]
  \includegraphics[width=1\textwidth]{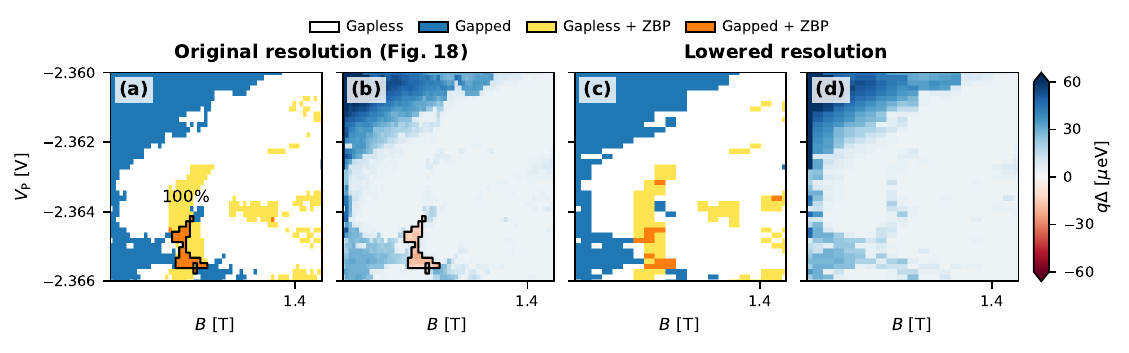}
  \vspace{-25pt}
	\caption{{\bf Dependence of the TGP outcome on resolution:} (i) TGP outcome for Device C (Fig.~18 of Ref.~\citenum{Agahee2023}) at the original resolution. This passes the TGP, with the orange region in (a) around $B\approx 1$~T identified as topological. (ii) The same device and measurement, but with a lower resolution (selecting every third pixel in $B$). Notably, this resolution is still similar to those used for other devices (see Fig.~\ref{pxs}). Despite this, the device now fails the TGP for this selection of resolution.}
	\vspace*{-0pt}
\label{resolution}
\end{figure*}

\renewcommand{\figurename}{Table.}
\setcounter{figure}{0}
\begin{figure*}[t]
	\centering
  \includegraphics[width=0.85\textwidth]{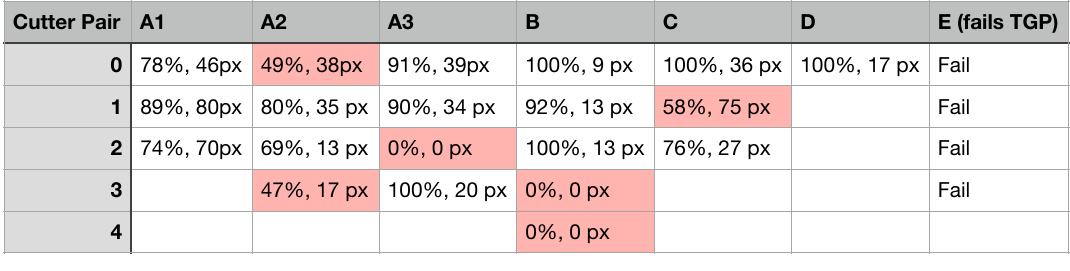}
	\caption{{\bf SOI$_2$s for different cutter pairs of the devices in Ref.~\citenum{Agahee2023}:} Gapless boundary percentage and number of pixels in SOI$_2$ from datasets for Ref.~\citenum{Agahee2023}. Other than Device A, measurement 1, which is the only comparison shown in Ref.~\citenum{Agahee2023}, all other devices with multiple cutters have a dependence on the chosen cutter pair and do not satisfy the TGP requirements for at least one SOI$_2$ (red entries). Even when they do satisfy the criteria, the SOI$_2$ can substantially differ in size for different cutter pairs (see, {\it e.g.}, sizes in A2). It should also be noted that there is a large variation in the number of cutter pairs for each experiment in Ref.~\citenum{Agahee2023}. The reason for the variations in cutter pair number is not explained.}	
\vspace*{-10pt}	
\label{cuttertable}
\end{figure*}

\renewcommand{\figurename}{Fig.}
\setcounter{figure}{5}
\begin{figure*}[htb]
  \includegraphics[width=1\textwidth]{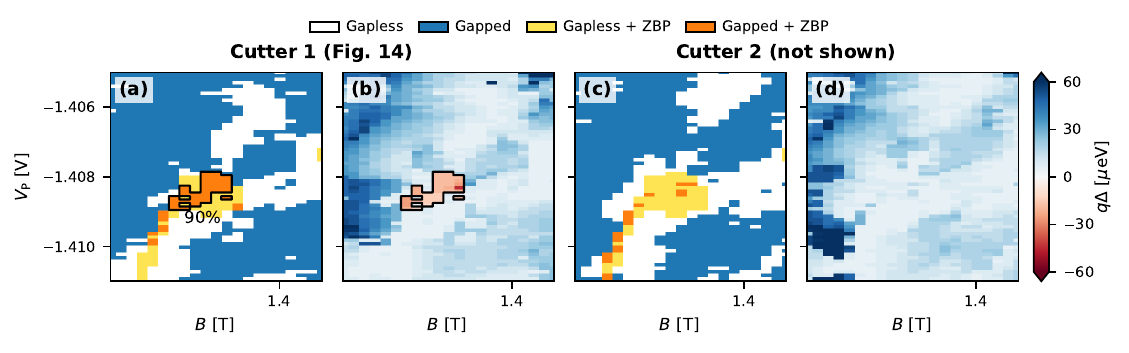}
  \vspace{-25pt}
	\caption{{\bf Dependence on cutter voltage pair:} {\bf (a-b)} TGP Subregion Of Interest 2 (SOI$_2$) for Device A3 using cutter pair 1 (Fig.~14 of Ref.~\citenum{Agahee2023}). This passes the TGP with the orange region around $B\approx 1$~T identified as topological. {\bf (c-d)} The same device and same measurement now with cutter pair index~2 shown. Although the overall magnitude of nonlocal conductance is largely unaffected by the change in cutter pair, the region identified around $B\approx 1$~T is now identified as gapless.}
\label{cutter}
\end{figure*}

{\bf (B2) Data resolution: Dependence and variations.}
In addition to variations in data ranges, the resolution of the data in Ref.~\citenum{Agahee2023} also differs significantly between measurements, even for the same device, see Fig.~\ref{pxs}. Such large changes in resolution are problematic because key aspects of the TGP code are defined in terms of pixel counts rather than fixed physical quantities. As a result, the  outcome of the TGP can be dependent on the experimental resolution choice.

A clear example of such a quantity in the TGP is the \textit{minimal cluster size} parameter. Namely, the TGP requires a region to contain at least 7 pixels to be identified as a region of interest (see code extract below). Since the size of a pixel --- both experimentally in units of mT $\times$ mV and physically in terms of $\mu$eV$^2$, based on the reported lever arms and g-factors in Ref.~\citenum{Agahee2023} --- varies considerably between different experiments (see Fig.~\ref{pxs}), the requirement of 7 pixels depends on the chosen resolution of the data. Increasing resolution (i.e., reducing pixel size) can cause a previously too-small region to {pass} the TGP, while decreasing resolution can merge separate pixels to form a continuous region, again altering the outcome. For example, in Fig.~\ref{resolution} reducing the resolution of Device C to a level comparable to other devices causes the region that previously passed the TGP (orange highlighted) to now fail the TGP.

We emphasise that there are several other thresholds within the TGP that are defined in terms of pixel numbers. For instance the position tolerance for distance between the end of the ZBP and the location of the purported gap. 

\vspace{5pt}
{\small
\begin{tcolorbox}[colback=gray!20, colframe=black, sharp corners, boxrule=0.5pt, width=\columnwidth]
\centering
\begin{verbatim}
756:	min_cluster_size: float | int = 7
\end{verbatim}
\end{tcolorbox}}
\noindent {\bf Code extract for setting minimal cluster size:} The minimal gapped cluster size is set to 7 pixels (extract from \href{https://github.com/microsoft/azure-quantum-tgp/blob/ed52b7a857d6d2f8a30bfbb19b16cf3b13d504f9/tgp/two.py}{two.py}). Since data resolution determines the number of pixels in a cluster, altering the resolution can change the TGP outcome.
\vspace{10pt}

{\bf (B3) Cutter pairs: Dependence and variations.}
In App.~F of Ref.~\citenum{Agahee2023} a comparison of the three different cutter pairs for Device A1 is shown and based on this it is stated: {\sl ``This comparison shows that SOI$_2$s} [Subregions of Interest 2] {\sl corresponding to the different cutter pairs are similar."} This statement gives the impression that it also holds for other devices. However, this is not the case. As shown in Table~\ref{cuttertable}, in all other devices with more than one cutter pair, at least one SOI$_2$ does not satisfy the TGP requirements (red boxes) and the size of the identified regions varies significantly. Furthermore, in several cases no SOI$_2$ for a given cutter pair is identified and the gapped region with ZBPs is identified as gapless. It should be emphasised that this occurs even when there is no appreciable change in the magnitude of the nonlocal conductance.
 
It is notable that the number of cutter pairs varies significantly across measurements (see Table~\ref{cuttertable}), from just one cutter in Device D  to five cutters in Device B. The reason for these different numbers of cutter pairs is not explained in Ref.~\citenum{Agahee2023} and all simulations have five cutter pairs. Perhaps surprisingly, since in Ref.~\citenum{Agahee2023} the SOI$_2$s are only required to satisfy the TGP conditions for 50\% of cutter pairs, all these devices still pass the TGP as set out in Ref.~\citenum{Agahee2023}. This also means that selection of various cutter pairs can alter the TGP result. For example, in Device B, choosing cutter pairs \{0,3,4\} would have resulted in the device now failing the TGP, even though this is the same number of cutter pairs as in Device A1 and Device C. Taken to the extreme, Device D has just one cutter pair. This means that virtually all devices can both pass and fail the TGP as applied to Device D, depending on a chosen individual cutter pair.

  \vspace*{0pt}
\section{Appendix C: Further discussion of how the TGP for theoretical simulations is not the same as that applied to experimental data.}
\vspace{0pt}
\begin{figure*}[t]
  \includegraphics[width=1\textwidth]{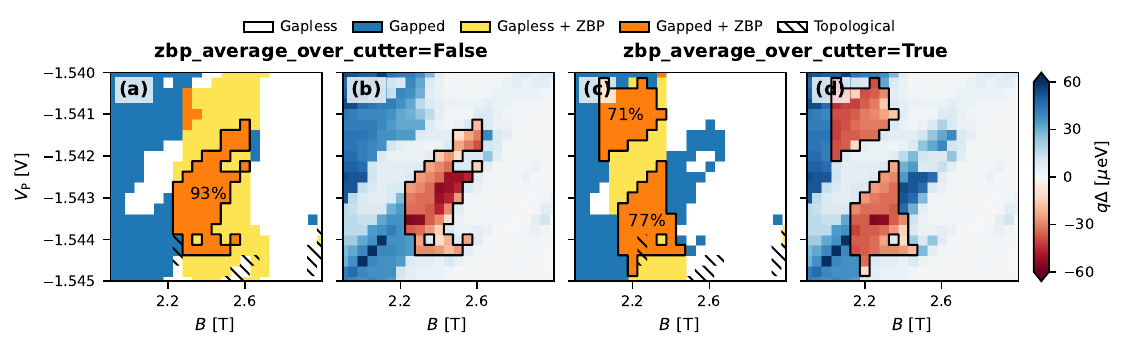}
  \vspace*{-25pt}
	\caption{{\bf False positive in experimental TGP applied to theoretical simulations:} {\bf (a-b)} Outcome of the TGP when average\_over\_cutter=False (as in \texttt{analyze\_2} applied to produce Table II of Ref.~\citenum{Agahee2023}) on simulated data file simulated\_DLG\_epsilon\_\_disorder\_seed\_5
\_\_geometry\_seed\_10\_\_surface\_charge\_4.0.nc. No false positive occurs for this parameter. {\bf (c-d)} Outcome of TGP as applied to experimental data (\texttt{analyze\_two}) on the same file, i.e., with average\_over\_cutter=True. There is a false positive region that passes the TGP for $V_{\rm p} \approx -1.541$~V. This shows that the claim of ``no false positives" is not accurate for the TGP code as actually applied to experimental data.}
	\vspace*{-5pt}
\label{2tgps}
\end{figure*}
Although the code for the simulations in Ref.~\citenum{Agahee2023} has not been released, applying the publicly available TGP code to the released simulation data allows us to examine the claim that the simulations contain ``no false positives".

As discussed in the main text, this claim is {\sl not} correct for the TGP as applied to experiments. For example, Fig.~\ref{2tgps}{\bf (c-d)} shows the experimental TGP applied to a dataset used to generate Table II of Ref.~\citenum{Agahee2023}. This portion of phase space has two identified regions that pass the TGP: one centred at ($B$,$V_{\rm p})=($2.3~T, $-$1.544~V$)$ and another centred at ($B$,$V_{\rm p})=($2.2~T, $-$1.541~V$)$. While the former is considered a true positive (despite mostly covering `trivial' phase space), the second region is a genuine false positive, even by the definitions utilised in Ref.~\citenum{Agahee2023}. This directly contradicts Table~II of Ref.~\citenum{Agahee2023}, which reports 0 false positives (FP) in every entry and forms the basis for the claim of a $<$8\% probability of a trivial device passing the protocol.

This discrepancy arises because the Python code used for Table~II (\href{https://github.com/microsoft/azure-quantum-tgp/blob/ed52b7a857d6d2f8a30bfbb19b16cf3b13d504f9/notebooks/yield_analysis.py}{\texttt{yield\_analysis.py}}) and the code used for the paper figures (\href{https://github.com/microsoft/azure-quantum-tgp/blob/ed52b7a857d6d2f8a30bfbb19b16cf3b13d504f9/notebooks/paper_figures.py}{\texttt{paper\_figures.py}}) utilise different implementations of the TGP. The former uses \texttt{analyze\_2} whilst the latter uses \texttt{analyze\_two} (see code excerpts below).

Crucially, the variable \texttt{average\_over\_cutter} --- which determines whether zero-bias peaks in the local conductance are averaged across different cutter voltage pairs --- is \texttt{False} in the simulation TGP but \texttt{True} in the experimental TGP. As Fig.~\ref{2tgps} demonstrates, this fundamentally alters the TGP outcome. The claim of zero false positives is therefore an artefact of using a different TGP for Table~II than for the experimental figures, demonstrating that there is no consistent definition of the protocol even within Ref.~\citenum{Agahee2023} itself.

\twocolumngrid

\renewcommand{\figurename}{Fig.}
\begin{figure*}[t]
	\centering
  \includegraphics[width=1\textwidth]{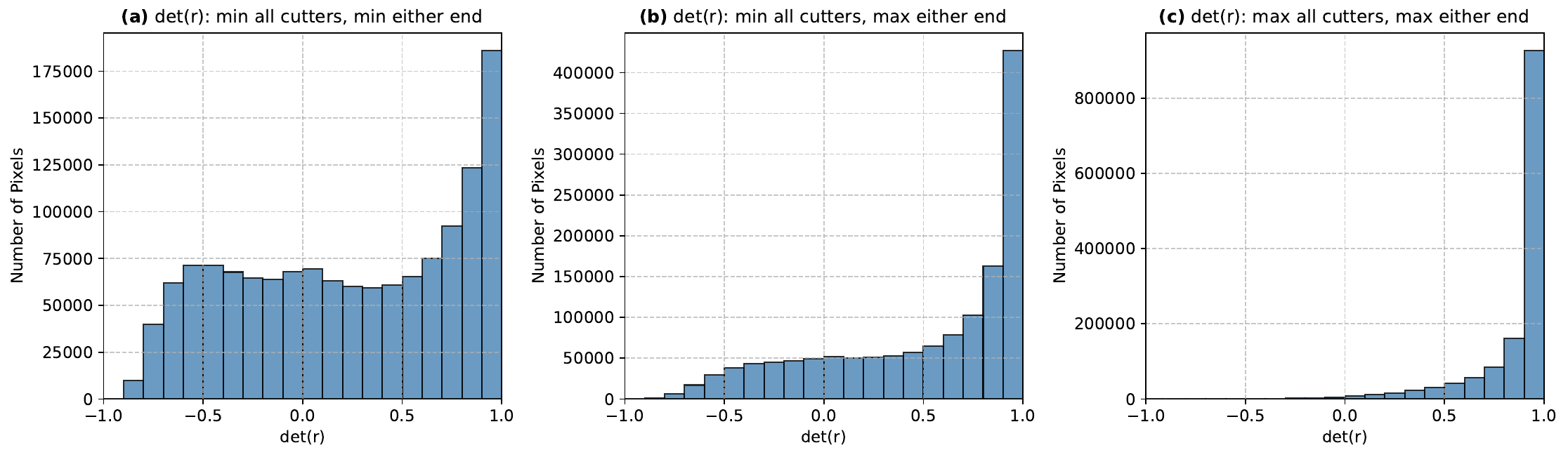}
	\caption{\textbf{Values of det($r$) in all simulations for Ref.~\citenum{Agahee2023} for various possible definitions of `topological'.} 
       Produced using data from all simulation files. {\sl Note changes in $y$-axis scale}. {\bf (a)} Here we show the  definition of topological pixels used in Ref.~\citenum{Agahee2023}, i.e., the minimum of $\det(r)$ for any cutters and minimum on either end of the nanowire. For this definition 37.7\% of all pixels in simulations are topological. Even here a vanishingly small number of pixels satisfy the Pikulin {\it et al.} definition of topology as $\det(r) < -0.9$. {\bf (b)} A slightly stronger definition would be to demand that, for any cutter, both ends contain $\det(r) < 0$. In this case the number falls to just 20.1\% of pixels satisfying the criterion. {\bf (c)} The most stringent criterion based on  $\det(r)$ would be to demand $\det(r) < 0$ on both ends for all cutters. We find just $0.9\%$ of pixels would satisfy this stringent criterion.}
\vspace{-10pt}
\label{histograms}
\end{figure*}

\begin{figure*}[t]
	\centering
  \includegraphics[width=1\textwidth]{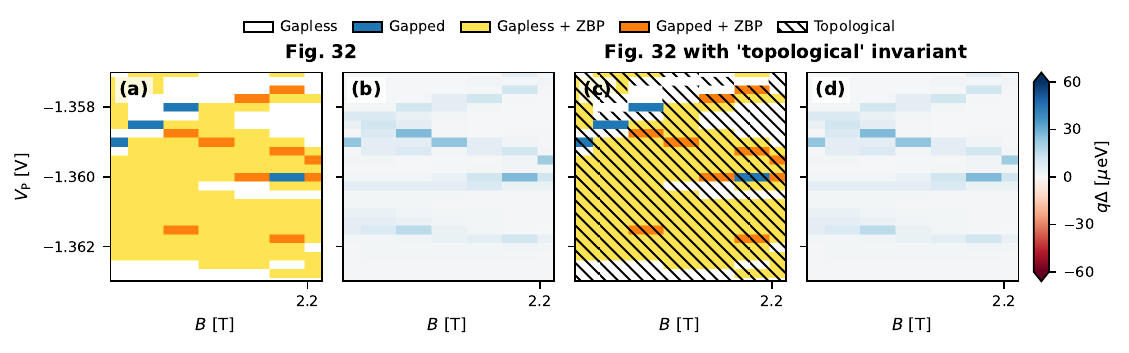}
    \vspace*{-25pt}
	\caption{\textbf{Inclusion of missing topological classification of pixels in the simulation failing the TGP.} 
        {\bf (a-b)} The original Fig.~32 as presented in Ref.~\citenum{Agahee2023}, showing gapped, gapless, and zero-bias peak (ZBP) regions in the simulated dataset. 
        {\bf (c-d)} The same figure with the additional overlay of topological classification of pixels (hashed regions), which were not shown in Ref.~\citenum{Agahee2023}. The omission of this classification in the original figure gives the impression that the simulated phase space is mostly trivial under the redefinition utilised by Ref.~\citenum{Agahee2023}, when, in fact, most of phase space is actually classified as topological. In contrast, in all other simulations the TGP was reported to pass and topological pixels were explicitly shown. The fact that almost all of phase space is classified as topological in this simulation raises questions about how topology is defined in Ref.~\citenum{Agahee2023}, but this was obscured by the omission of topological classification of pixels.
       } 
 \vspace{-15pt}
\label{simulation}
\end{figure*}

\section{Appendix D: Further discussion of the redefinition of topological.}
As discussed in the main text, to classify a simulated pixel as topological, the TGP uses the determinant of the reflection matrix evaluated at zero-bias, $\det(r)$~\cite{Fulga2011}. The original TGP defined by Pikulin \textit{et al.}~\cite{Pikulin2021} required a stringent scattering invariant of $\det(r) < -0.9$. In contrast, Ref.~\citenum{Agahee2023} significantly weakens this criterion, classifying a pixel as topological if $\det(r) < 0$ at \textsl{either} nanowire end for \textsl{any} cutter pair (the ``union'' of $\det(r)$).

This weakened definition classifies 37.7\% of the entire simulated phase space as topological. Ref.~\citenum{Agahee2023} claims this topological index is ``relatively insensitive to these details of the junctions'', but Fig.~\ref{histograms} demonstrates otherwise. Demanding $\det(r) < 0$ at both left and right junctions for at least one cutter pair drops the topological phase space to 20.1\%. Applying the most stringent criterion --- $\det(r) < 0$ at both ends for \textsl{all} cutters --- reduces it to just 0.9\%. The definition of topology in Ref.~\citenum{Agahee2023} is therefore highly dependent on junction details. As discussed in the main text, the extent of phase space that was classified as topological was obscured by selective presentation. In particular, Fig.~\ref{simulation} demonstrates that almost all of phase space was classified as topological in Fig. 32 of Ref.~\citenum{Agahee2023}.

The criteria for identifying an entire region as a ``true positive'' are even weaker. An identified region (minimum 7 pixels) is deemed a true positive if it overlaps with just a single topological pixel. With 37.7\% of the phase space classified as topological, the probability of random overlap is exceptionally high. For example, the region in Fig.~\ref{2tgps} centred at $(B, V_{\mathrm{p}}) = (2.2~\mathrm{T}, -1.541~\mathrm{V})$ is designated a true positive based on an overlap with only a few topological pixels.

Consequently, the true positives reported in the simulations of Ref.~\citenum{Agahee2023} can be entirely unrelated to Majorana zero modes, and the central claim that devices passed the protocol ``defined in Pikulin \textit{et al.}'' is incorrect.
\vspace{-10pt}

\end{document}